\documentclass[prl,floatfix,superscriptaddress,amsmath,amssymb,twocolumn]{revtex4}
\usepackage[latin1]{inputenc}
\usepackage{amsmath}
\usepackage{amssymb}
\usepackage{graphicx}
\usepackage{graphics}
\usepackage{color}
\usepackage{multirow}
\DeclareMathAlphabet{\mathit}{OT1}{ptm}{m}{it}

\usepackage{version}

\usepackage{subfigure}

\def \ba {\begin{eqnarray}}
\def \ea {\end{eqnarray}}

\newcommand{\nn}{\nonumber}
\newcommand{\ket}[1]{| #1 \rangle}
\newcommand{\bra}[1]{\langle #1 |}

\newcommand{\op}[1]{\mathbf{#1}}

\newcommand{\ta}{\Theta_{AF}}

\newcommand{\sew}{\mathit{w}}

\newcommand \bd  {\begin{details}}
\newcommand \ed  {\end{details}}

\begin{document}

\excludeversion{details}

\title{$Z_2$ antiferromagnetic topological insulators with broken $C_4$ symmetry }

\author{Fr\'{e}d\'{e}ric B\`{e}gue} \affiliation{Laboratoire de Physique Th\'eorique,
IRSAMC, CNRS and Universit\'e de Toulouse, UPS, F-31062 Toulouse,France}

\author{Pierre Pujol} \affiliation{Laboratoire de Physique Th\'eorique,
IRSAMC, CNRS and Universit\'e de Toulouse, UPS, F-31062 Toulouse,France}

\author{Revaz Ramazashvili} \affiliation{Laboratoire de Physique Th\'eorique,
IRSAMC, CNRS and Universit\'e de Toulouse, UPS, F-31062 Toulouse,France}


\begin{abstract}
A two-dimensional topological insulator may arise in a centrosymmetric 
commensurate N\'{e}el antiferromagnet (AF), where staggered magnetization 
breaks both the elementary translation and time reversal, but retains their product 
as a symmetry. Fang {\it et al.},~[Phys. Rev. B {\bf88}, 085406 (2013)] proposed 
an expression for a $Z_2$ topological invariant to characterize such systems. 
Here, we show that this expression does not allow to detect all the existing 
phases if a certain lattice symmetry is lacking. 
We implement numerical techniques to diagnose topological phases 
of a toy Hamiltonian, and verify our results by computing the Chern 
numbers of degenerate bands, and also by explicitly constructing 
the edge states, thus illustrating the efficiency of the method.
\end{abstract}


\maketitle

Physical phenomena, whose description involves topology, have been invariably 
attracting attention regardless of whether the word ``topology" was actually used 
at the time: early examples involve topologically non-trivial stable defects such as dislocations in crystals, as well as vortices in superconductors 
and superfluids. Quantum Hall Effect and its remarkably precise conductance 
quantization~\cite{cage2012quantum}
marked the advent of an entirely new class of phenomena, related not so much 
to the appearance in the sample of finite-size topological objects, but rather to 
the electron state of the {\em entire} sample changing its topology in a way, 
that could no longer be undone by a local perturbation. More recently, it was 
understood that, in fact, non-trivial topology may appear even in zero 
magnetic field: the fact that a commonplace band insulator may find itself 
in distinct electron states that cannot be continuously transformed one into 
another without a phase transition, came as a major 
surprise~\cite{ReviewHasanKane,ReviewQiZhang}.  

These phenomena invite the question of classifying topologically distinct 
states of matter: labeling each state by a set of discrete indices in such 
a way as to have different sets for any two phases that cannot be 
continuously transformed one into another without the system undergoing 
a phase transition. In the general setting, the problem remains unsolved. 

In fact, open questions are present even in a non-interacting description of 
systems that are believed to admit a $Z_2$ (``even-odd") classification, and 
thus have only one topologically trivial and one topologically non-trivial phase, 
commonly called topological. Here, we address one such question, that has 
recently attracted attention: diagnosing the topological phase of a $Z_2$ 
insulating N\'eel antiferromagnet. 

To put the subsequent presentation in context, we recapitulate the key results 
for the prototypical $Z_2$ system: a paramagnetic topological insulator. 
Fu and Kane~\cite{Fu_Kane2007} have shown that the $Z_2$ invariant 
for such a system can be defined via the so-called sewing matrix 
$\sew(\op{k})_{mn}$: 
\begin{equation}
\sew(\op{k})_{mn}=\bra{\Psi_{m,\op{-k}}} \Theta\ket{\Psi_{n,\op{k}}} ,
\end{equation} 
where the $\ket{\Psi_{n,\op{k}}}$ is the Bloch eigenstate of the $n$-th band 
at momentum $\op{k}$. The $\sew(\op{k})_{mn}$ turns out to be of particular 
interest at special momenta $\Gamma_i$ such that 
$-\Gamma_i=\Gamma_i+\op{G}$, with $\op{G}$ a reciprocal lattice vector. 
Such $\Gamma_i$ are now commonly called the ``time reversal-invariant momenta" 
(TRIM). In the Brillouin zone, a $\Gamma_i$ is  equivalent to its opposite, and thus 
the $\sew (\Gamma_i)_{mn}$ is antisymmetric. Since each band has its Kramers 
partner, the number of bands at hand is even, and the above two properties allow 
one to define the Pfaffian $Pf [\sew(\Gamma_i)_{mn}]$. 
As established by Fu and Kane~\cite{Fu_Kane2007}, the $Z_2$ topological 
invariant $\Delta$ can be expressed in a continuous gauge via the $\sew(\Gamma_i)_{mn}$ as per 
\begin{equation} 
\label{eq.inv_topo_sew}
(-1)^\Delta = \prod_{i} \frac{\sqrt{\det[\sew (\Gamma_i)]}}{Pf [\sew (\Gamma_i)]} . 
\end{equation} 
Moreover, in the presence of inversion symmetry, the Eq. (\ref{eq.inv_topo_sew}) 
may be recast in terms of the parity eigenvalues $\xi(\Gamma_i)$ of Bloch 
eigenstates at the TRIM $\Gamma_i$:
\begin{equation}\label{eq.inv_topo_xi}
(-1)^\Delta=\prod_{i} \prod_{\alpha}^N \xi_{\alpha}(\Gamma_i)
\end{equation}
where the $i$ labels the TRIM and the $\alpha$ 
counts one band per each pair of the $2N$ filled Kramers-partner bands.

N\'eel antiferromagnetism explicitly violates the symmetry with respect to time 
reversal $\Theta$. However, both the $\Theta$ and the translation ${\bf T_a}$ 
by half the N\'eel period invert the local magnetization, and thus the combination 
$\Theta_{AF} \equiv \Theta {\bf T_a}$ of the two remains a symmetry.

A number of authors \cite{MongEssinMoore,Guo_Feng_Shen,Fang_Gilbert_Bernevig,Liu14,Zhang15,Fang15,Liu13AF} 
attempted to classify the topological phases that may appear 
in an antiferromagnet. However, contrary to the paramagnetic case, 
the relevant anti-unitary operator does not square to -1: instead, its 
action on a Bloch eigenstate $\ket{\Psi_{n,\op{k}}}$ is given by 
\begin{equation}
\ta^2 \ket{\Psi_{n,\op{k}}} =-e^{i2\op{k}.\op{a}} \ket{\Psi_{n,\op{k}}}.
\end{equation} 
In the presence of inversion symmetry ${\bf I}$, the combined 
symmetry ${\bf I} \Theta_{AF}$ enforces double degeneracy at all momenta in the 
Brillouin zone. Moreover, the TRIM split into two kinds: the A-TRIM, 
where $\Theta_{AF}^2=1$ -- and the B-TRIM, where $\Theta_{AF}^2=-1$. 
In three dimensions, it has been argued that the B-TRIM suffice
to define a topological invariant, as the Eq. (\ref{eq.inv_topo_sew}) remains 
gauge-invariant as long as the product in the  r.h.s. is taken over the B-TRIM 
only \citep{Fang_Gilbert_Bernevig,Zhang15,Liu13AF}. Similarly, Fang {\it et al.} 
argued that in two dimensions, the product of the parity eigenvalues 
$\xi$ at the two B-TRIM would also be a $Z_2$ topological invariant.

The expression above, restricted to the B-TRIM only, appeared to work in the 
cases studied in the Refs. \citep{Fang_Gilbert_Bernevig,Zhang15,Liu13AF,Nous}. 
Such an expression tacitly implies, that the parity eigenvalues at the two A-TRIM 
either change simultaneously or not at all, and thus do not affect the $Z_2$ invariant. 
However, if a band inversion were to occur only at a {\em single} A-TRIM, then 
the full $Z_2$ invariant of the Eq. (\ref{eq.inv_topo_xi}) would change sign, and 
this would {\em not} be accounted for by the invariant, involving the parity 
eigenvalues at the B-TRIM only. 

Below, we use a method developed in a previous 
work~\cite{Nous} to illustrate this possibility by a toy Hamiltonian. 

{\it i) The model --} We consider a non-interacting electron system on a square lattice 
(of lattice vectors $a\op{\hat{X}}$ and $a\op{\hat{Y}}$), with an s- 
and a p-wave symmetry orbital on each site, as in the Bernevig-Hughes-Zhang (BHZ) 
model~\cite{BHZ}. The lattice can de divided into two sub-lattices, A and B 
(see Fig.(\ref{fig.Lattice})), corresponding to the opposite orientation of magnetization 
in the $z$ direction. In this case, the natural lattice vectors for the 
super-lattice will be $a\sqrt{2}\op{\hat{x}}=a\op{\hat{X}}+a\op{\hat{Y}}$ 
and $a\sqrt{2}\op{\hat{y}}=-a\op{\hat{X}}+a\op{\hat{Y}}$. In what follows, 
we choose $a=\frac{1}{\sqrt{2}}$. In this case, the TRIM will correspond 
to $(k_x,k_y)=(0,0),(\pi,\pi),(0,\pi)$ and $(\pi,0)$, the first two being 
the B-TRIM and the latter two being the A-TRIM. The Bloch Hamiltonian 
$H(\op{k}) = e^{-i\op{k}\op{r}} H e^{i\op{k}\op{r}}$ can be written as:

\begin{figure}
\includegraphics[height=5cm]{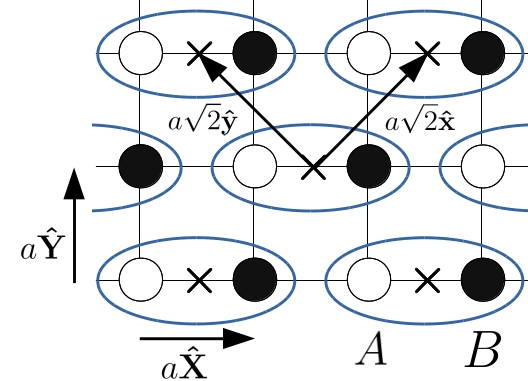} 
\caption{Square lattice, on which the Hamiltonian is defined. In the absence of a staggered magnetic field, the primitive Bravais lattice vectors are $a\op{\hat{X}}$ and $a\op{\hat{Y}}$. In the presence of staggered magnetization, the dimerized lattice is defined by the 
primitive vectors $a\sqrt{2}\op{\hat{x}}$ and $a\sqrt{2}\op{\hat{y}}$. In this case, a unit 
cell (in blue) comprises two sites, A and B (white and black dots, respectively).}
\label{fig.Lattice}
\end{figure}

\begin{align}\label{eq.BHZ+mag k space}
H(\op{k})&=\mu+\Delta\mu \tau^z \nn\\
&-2t(C_-+C_+) (C_-\sigma^x+S_-\sigma^y) \nn\\
&-2(t'_x \cos(k_x)+t'_y \cos(k_y))\tau^z \nn\\
&+2\alpha ( S_+ C_+ s^y\tau^y \sigma^z -S_- C_-  s^x\tau^y \sigma^z )\nn\\
&+m s^z \sigma^z 
\end{align}
where $C_\pm \equiv \cos\left[(k_x \pm k_y)/2\right]$ and 
$S_\pm \equiv \sin\left[(k_x \pm k_y)/2\right]$, while $\sigma$, $s$ 
and $\tau$ are the Pauli matrices acting in the sub-lattice (A and B), 
spin and orbital spaces, respectively.
The first term ($\mu_\pm=\mu\pm\Delta\mu$) originates from the energy 
difference of the the s- and p-symmetric orbitals. The second term corresponds 
to the nearest-neighbor hopping between the same orbitals, the third -- to 
second-neighbor hopping. We choose the latter to be anisotropic and 
orbital-dependent. This term breaks the $C_4$ symmetry, as explained later. 
The following term hybridizes the two orbitals via the amplitude $\alpha$, and is 
of a spin-orbital nature, it is a third nearest neighbor hopping. This term is 
responsible for a gap at half-filling, and thus for bulk insulating behavior. 
Finally, the last term describes the staggered magnetization.

In the following, we choose $\mu=0, \Delta\mu=3, t=1, t'_x=1, t'_y=0.5, \alpha=2$ 
and $m>0$. This choice is made to render the figures more legible, the same 
conclusions hold for more realistic parameters, such that $\alpha<t'_x,t'_y<t$.

Upon variation of $m$, the criterion due to Fang {\it et al.} would predict a single 
phase transition at $m=6$. For $0<m<6$, the product of the parity eigenvalues 
over a half of the filled bands (for every pair of doubly degenerate bands, 
such a product counts only a single parity eigenvalue)  
at the B-TRIM equals $-1$, and thus the system should be in a topological phase, 
while for $m>6$ this product is equal to $1$, and so the system should be in the 
trivial phase. 

However, this criterion tacitly assumes that no topological phase transition may 
take place via closing the gap at an A-TRIM. Indeed, this is true if the system 
is $C_4$-symmetric: this symmetry would guarantee that band inversion could 
occur only at both of the A-TRIM simultaneously, thus keeping the topological 
invariant intact. However, our Hamiltonian explicitly breaks the $C_4$ symmetry 
via the terms $t'_x$ and $t'_y$, hence the argument above no longer applies. 

\begin{figure}
\includegraphics[height=1.5cm]{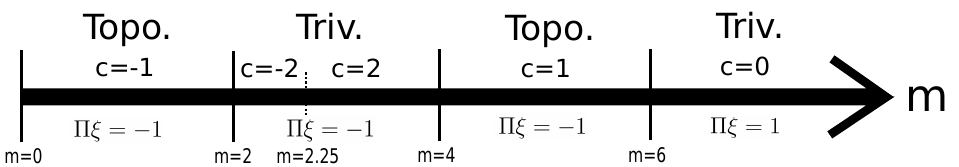} 
\caption{The phase diagram, obtained by the WCC method for 
$\mu=0, \Delta\mu=3, t=1, t'_x=1, t'_y=0.5, \alpha=2$ and $m>0$. 
``Topo.'' corresponds to the topological phase (an odd jump 
of the WCC position), and ``Triv.'' to the trivial phase (an even jump). 
Below we show the total Chern number of the two lowest energy bands, 
restricted to the stable subspace spanned by the four states 
$(\ket{\uparrow~sA},\ket{\uparrow~sB},\ket{\downarrow~pA},\ket{\downarrow~pB})$. 
The $\Pi\xi$ corresponds to the product of half the parity eigenvalues of the Bloch 
states at the B-TRIM. For $2<m<4$, the phase is trivial while $\Pi\xi=-1$, 
showing that $\Pi\xi$ alone cannot characterize the topology of the phase.}
\label{fig.DiagPhase}
\end{figure}

\begin{figure*}
\subfigure[$m=1$]{\includegraphics[trim = 0mm 38mm 0mm 39mm, clip,height=2.7cm]{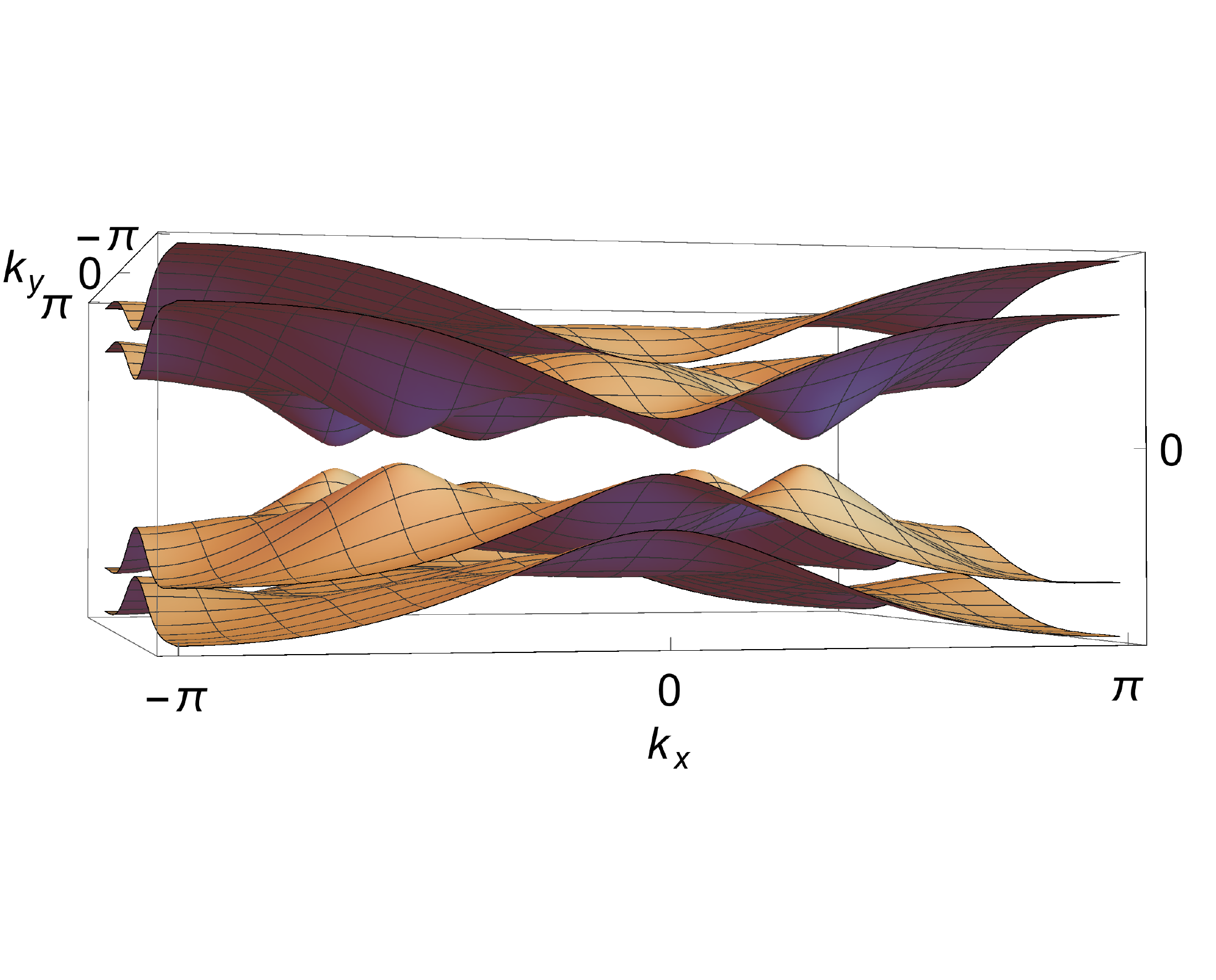}}
\subfigure[$m=2$]{\includegraphics[trim = 0mm 38mm 0mm 39mm, clip,height=2.7cm]{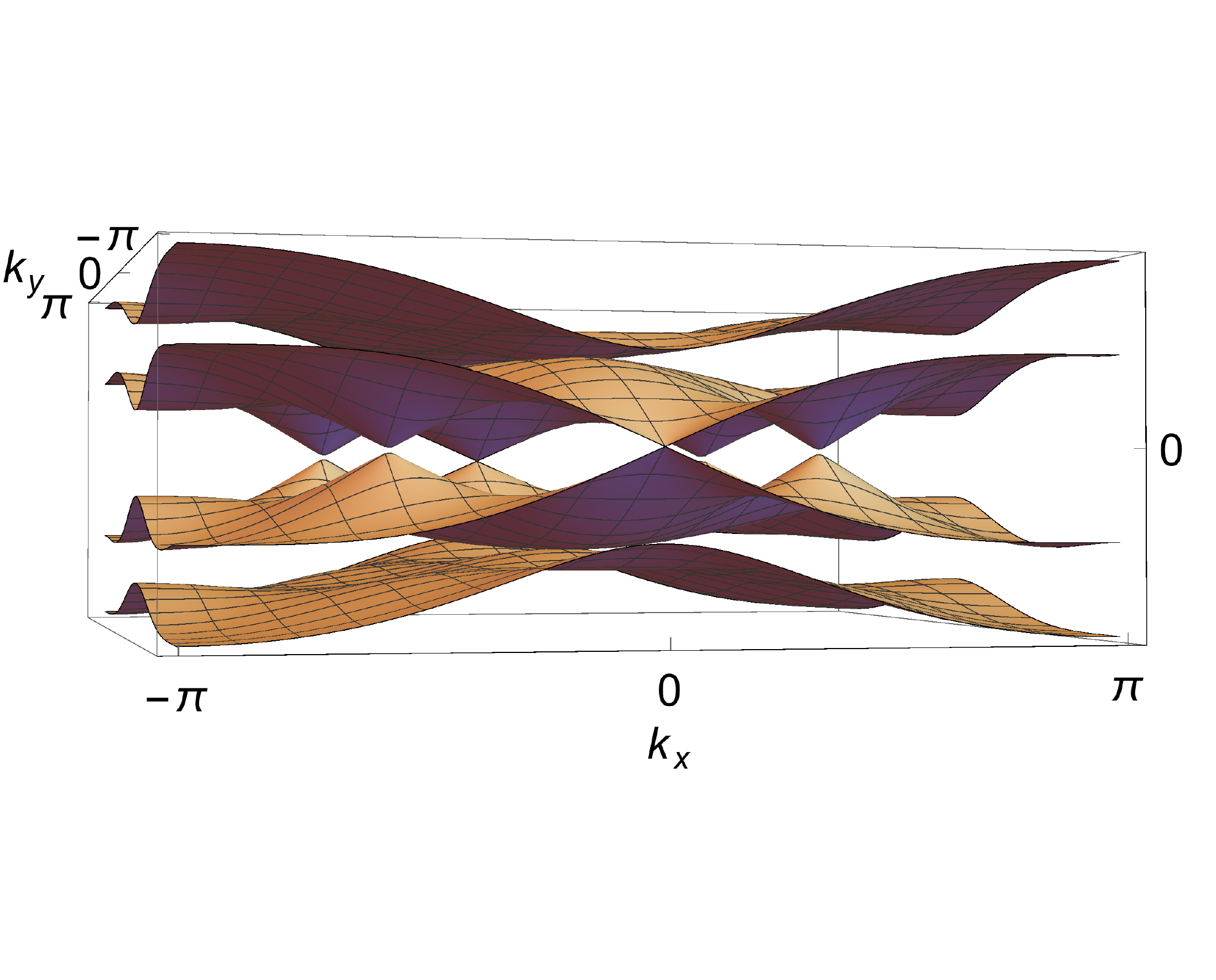}}
\subfigure[$m=2.25$]{\includegraphics[trim = 0mm 38mm 0mm 39mm, clip,height=2.7cm]{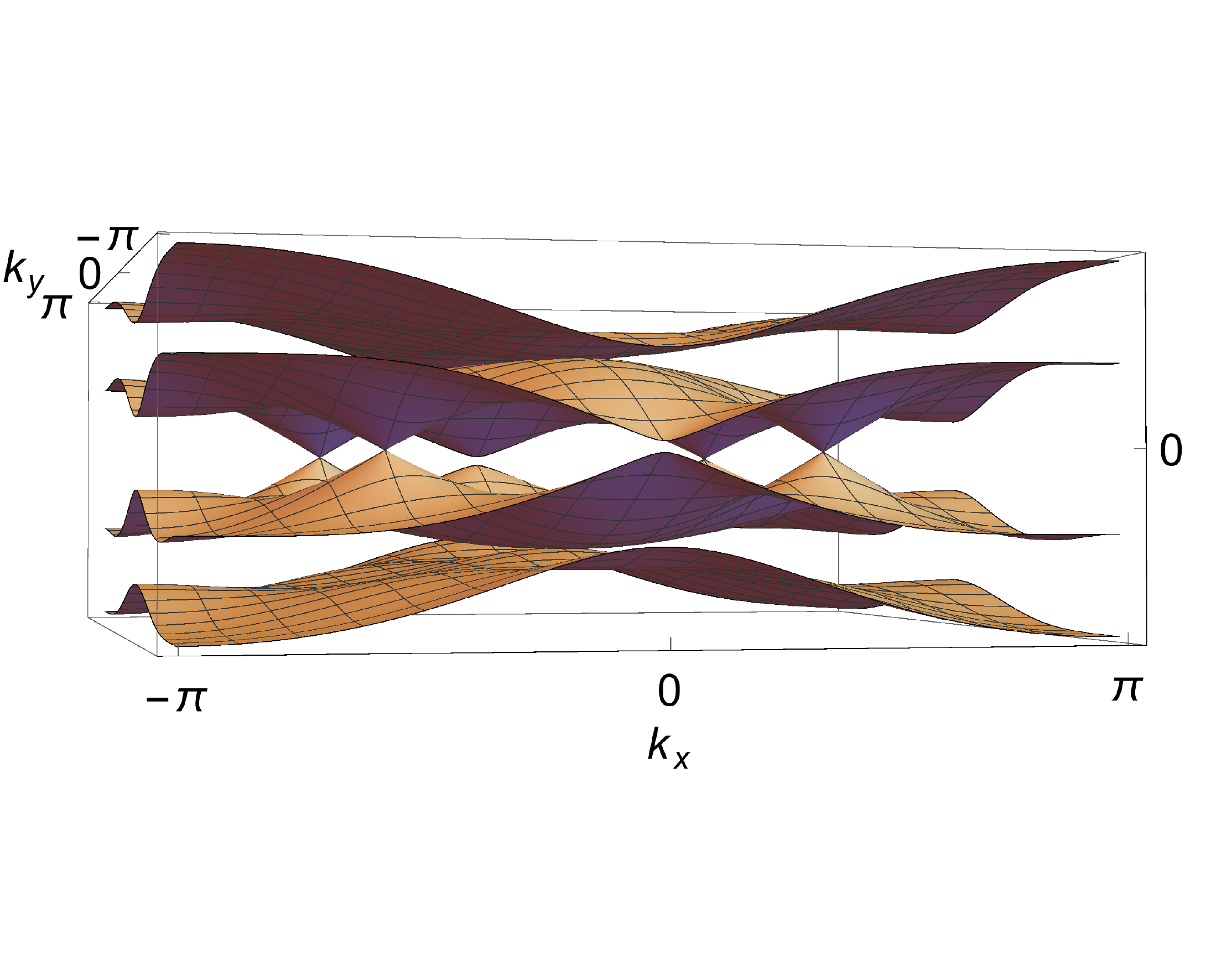}}

\subfigure[$m=3$]{\includegraphics[trim = 0mm 38mm 0mm 39mm, clip,height=2.7cm]{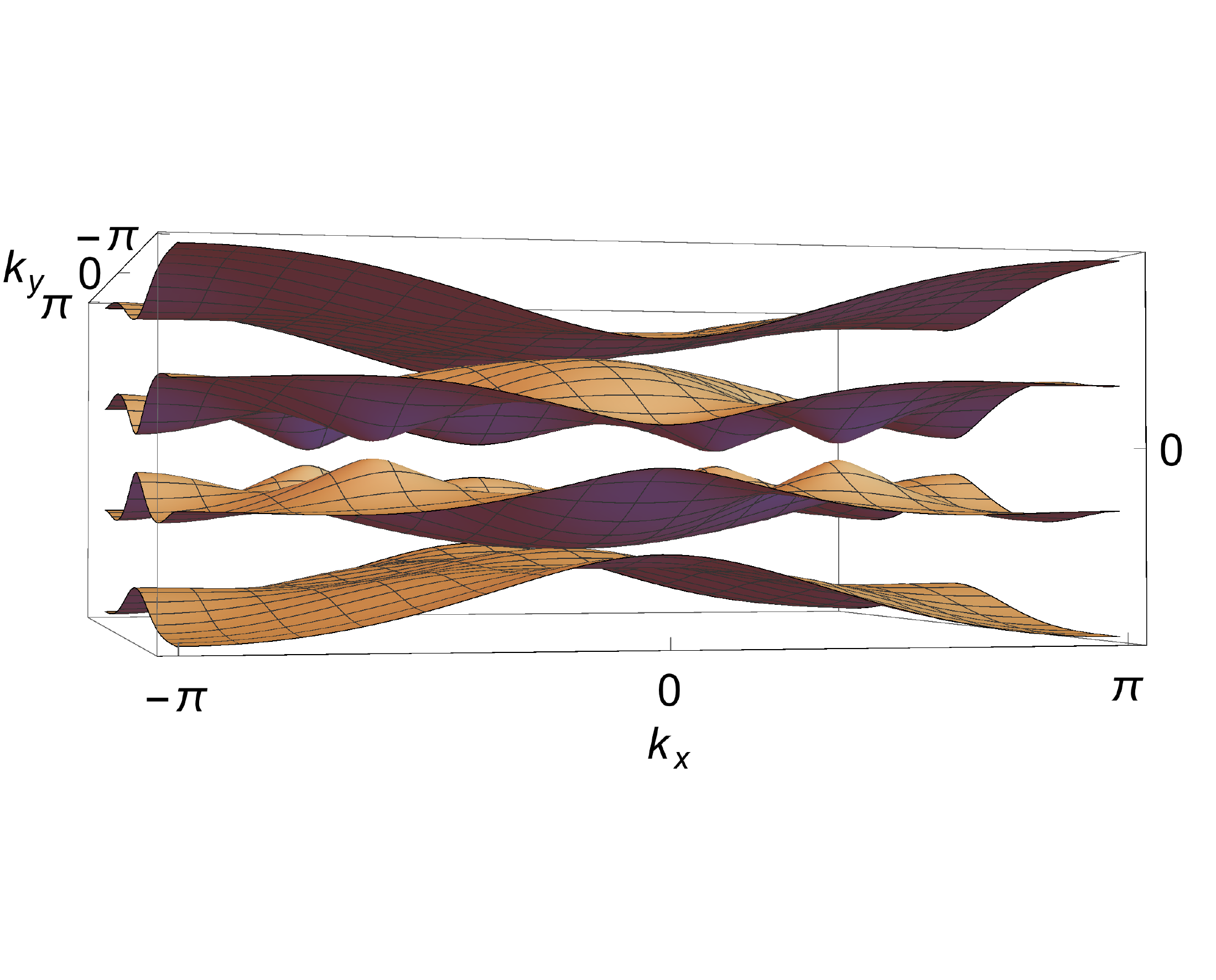}}
\subfigure[$m=4$]{\includegraphics[trim = 0mm 38mm 0mm 39mm, clip,height=2.7cm]{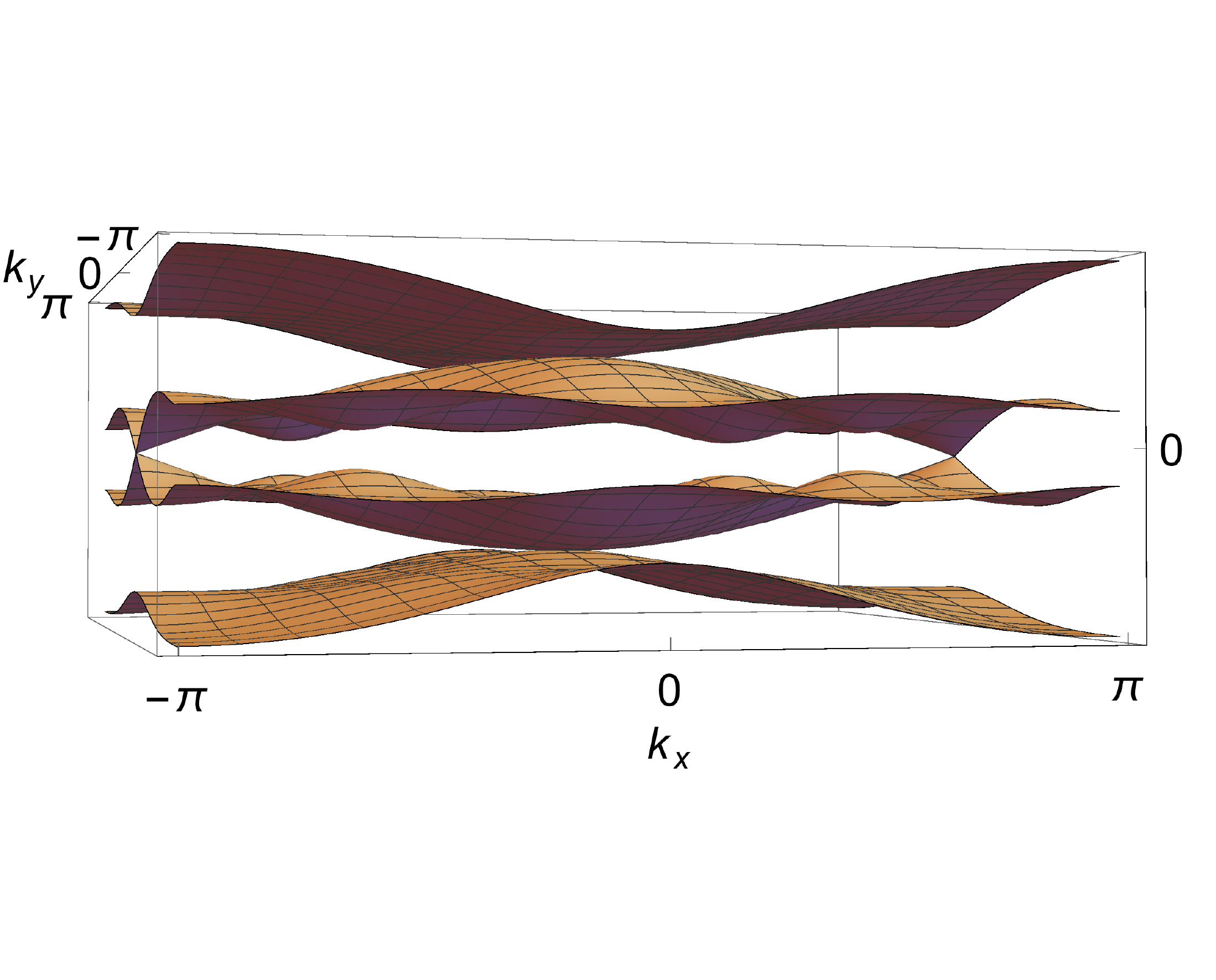}}
\subfigure[$m=5$]{\includegraphics[trim = 0mm 38mm 0mm 39mm, clip,height=2.7cm]{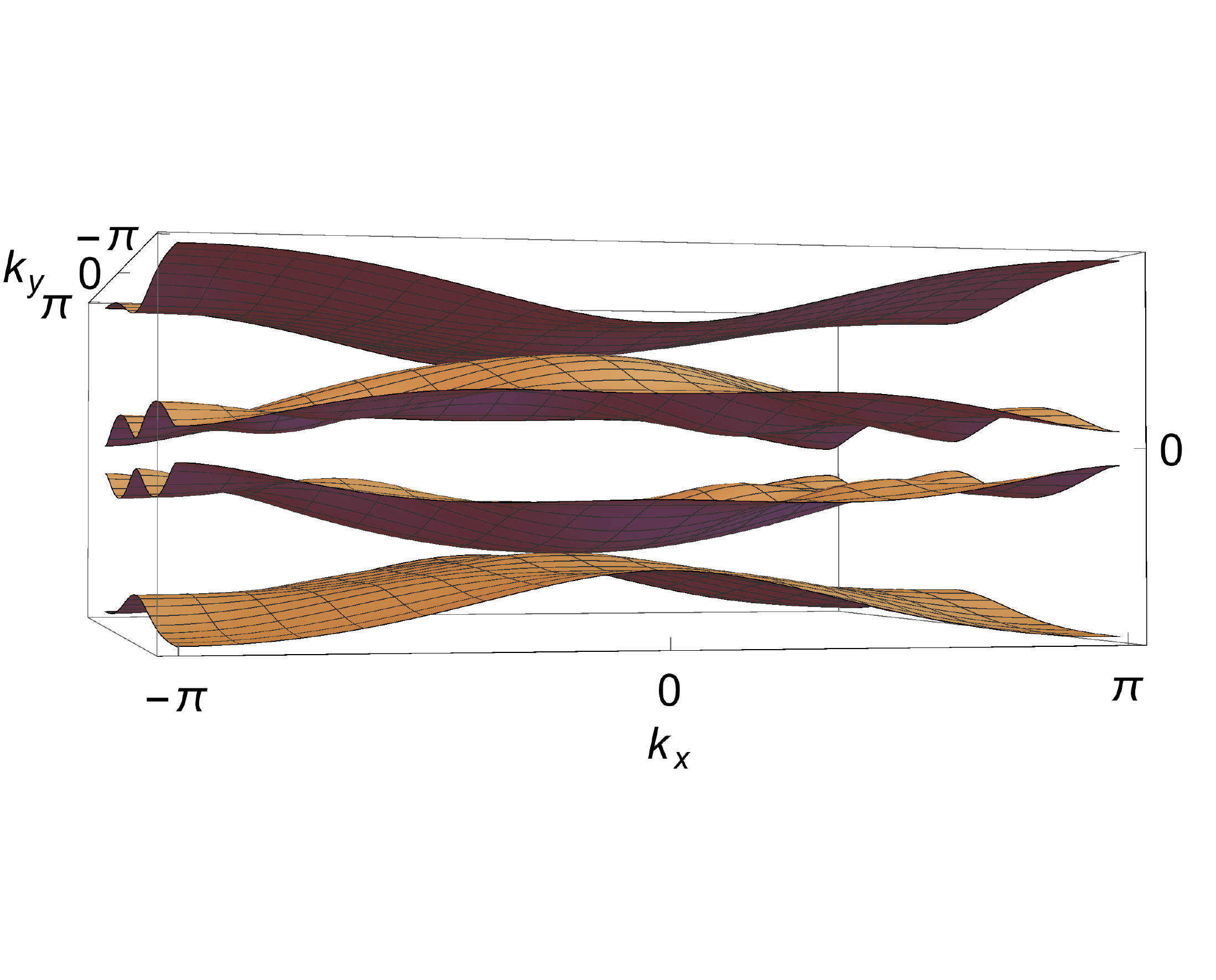}}

\subfigure[$m=6$]{\includegraphics[trim = 0mm 38mm 0mm 39mm, clip,height=2.7cm]{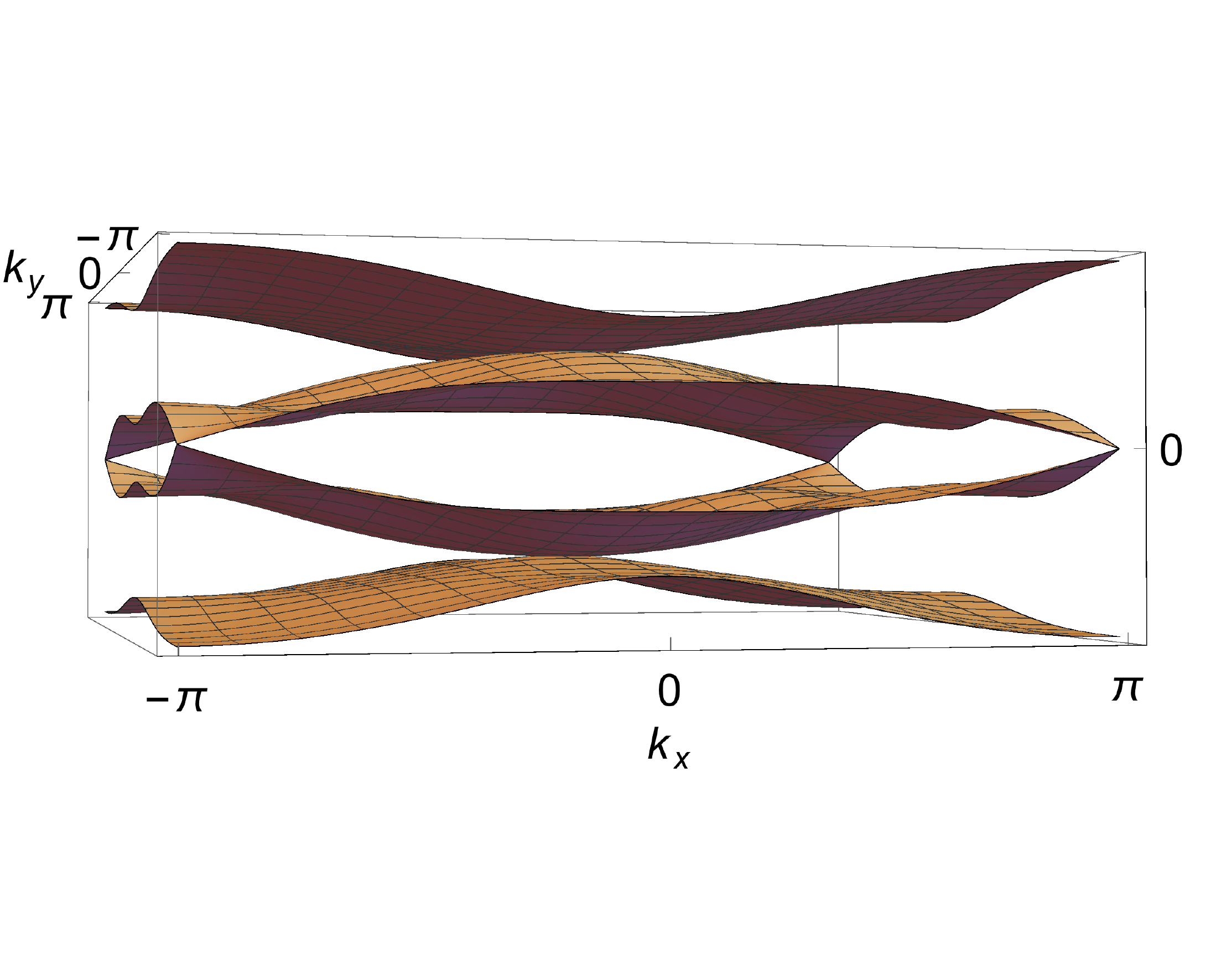}}
\subfigure[$m=7$]{\includegraphics[trim = 0mm 38mm 0mm 39mm, clip,height=2.7cm]{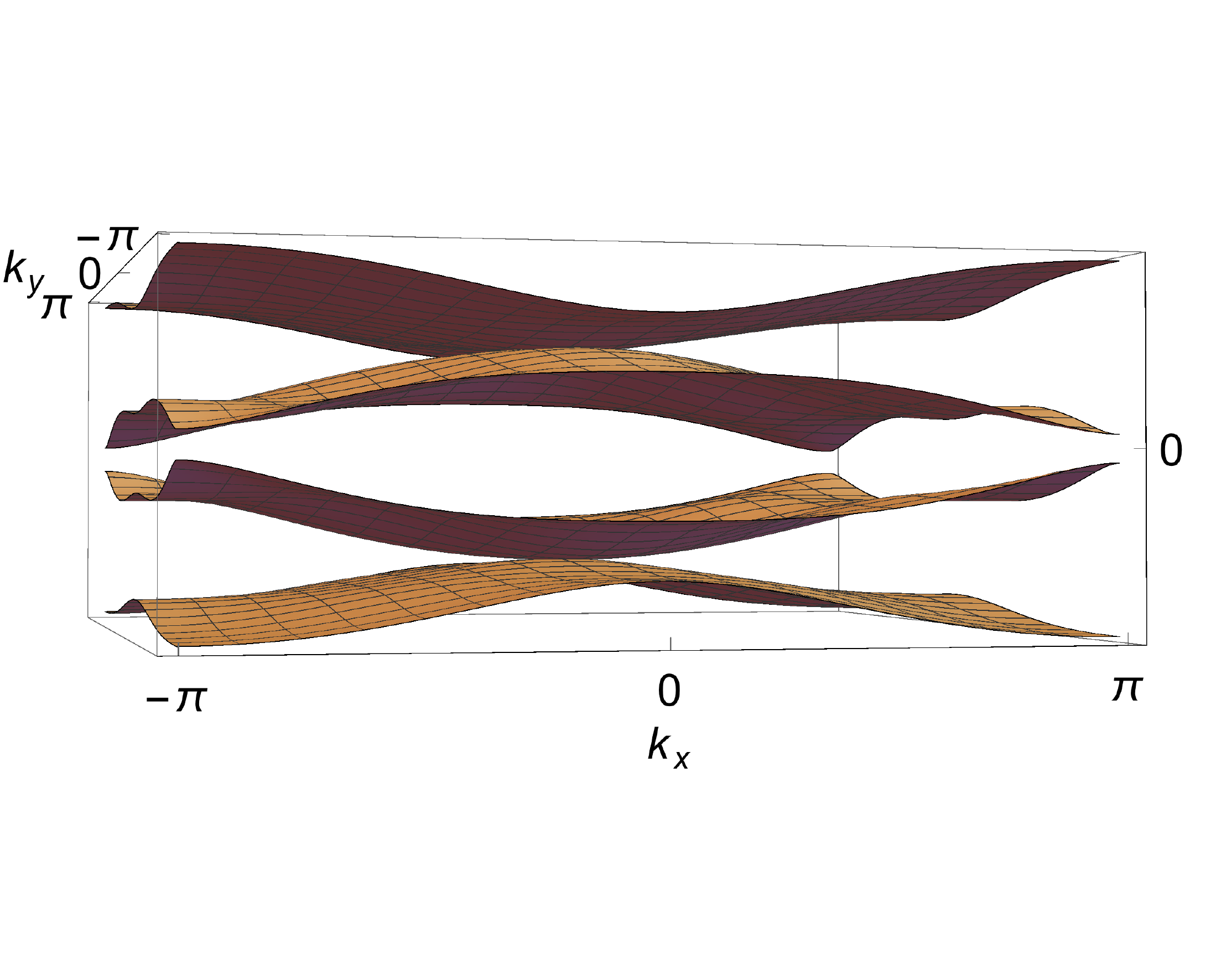}}

\caption{Dispersion relation for $\mu=0, \Delta\mu=3, t=1, t'_x=1, t'_y=0.5, \alpha=2$ and different values of $m$. The gap closes at a B-TRIM for $m=6$ and at the A-TRIM for two different values of m (2 and 4). For other values of m, a gap is present all over the BZ, insuring an insulating phase, except at $m\simeq 2.25$ where the gap closes at non TRIM points.}
\label{fig.EqDisp}
\end{figure*}

\begin{figure*}
\includegraphics[height=3cm]{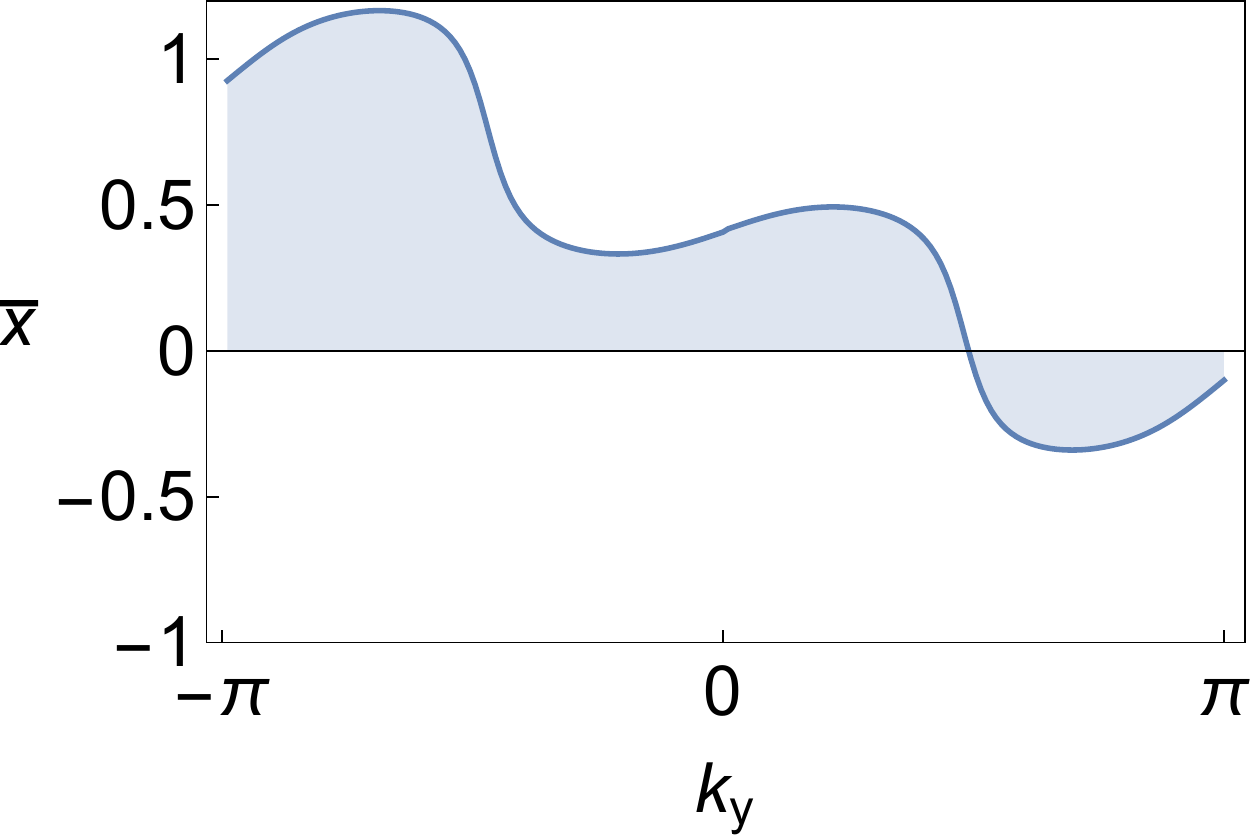} \includegraphics[height=3cm]{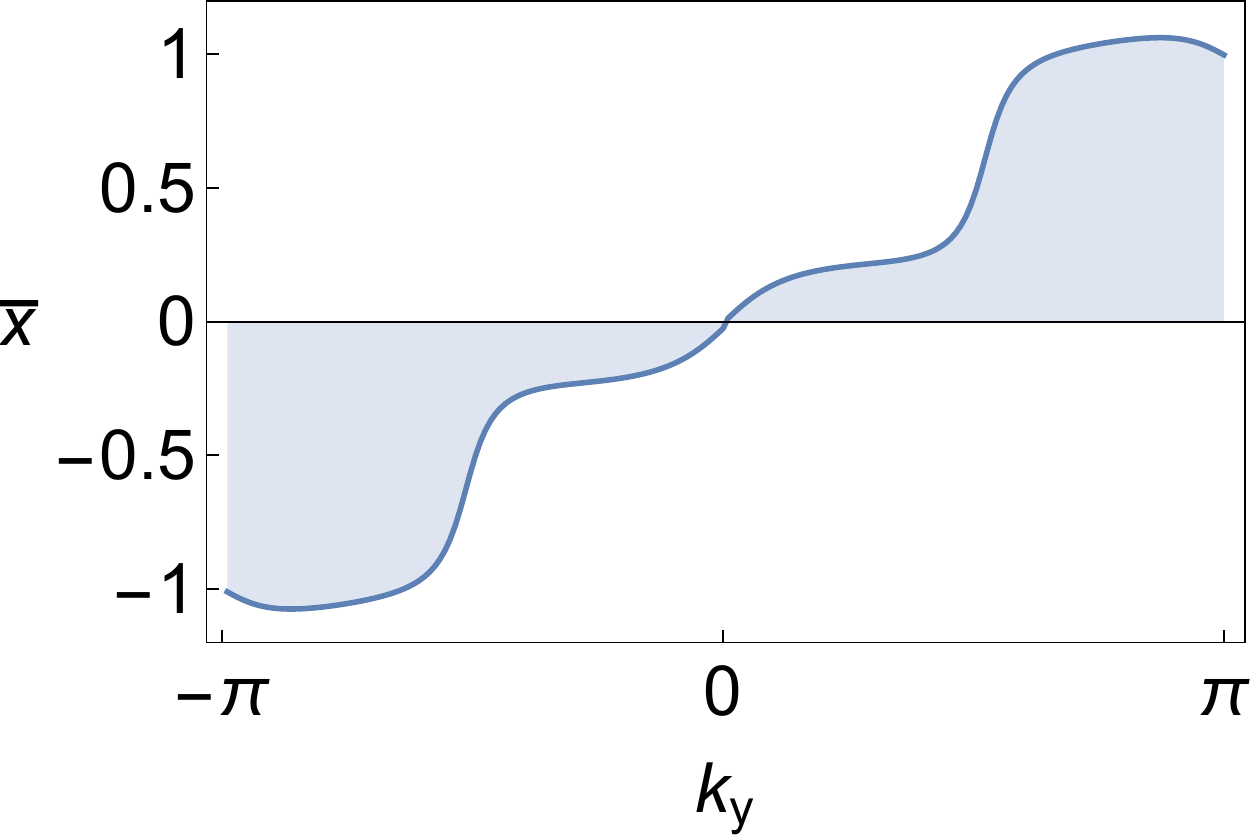} \includegraphics[height=3cm]{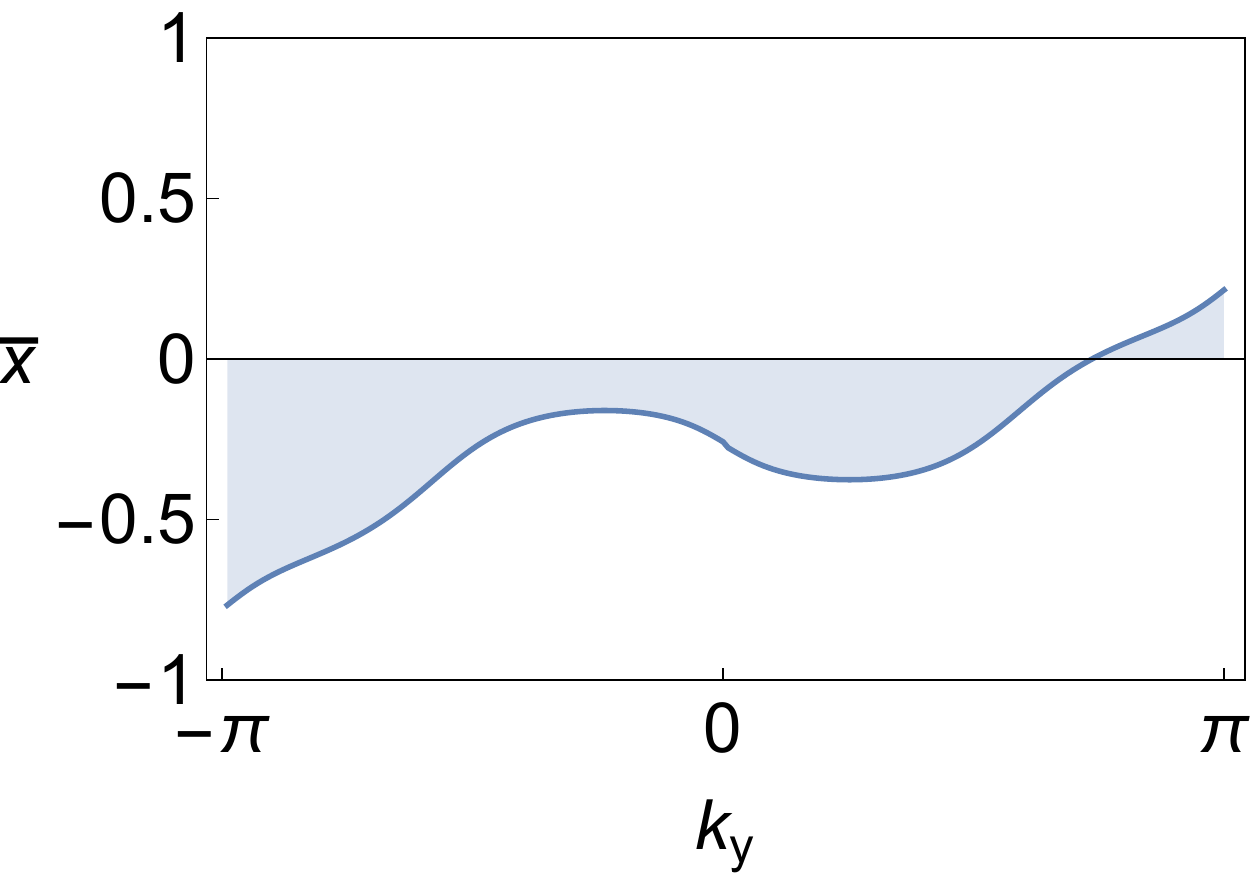} \includegraphics[height=3cm]{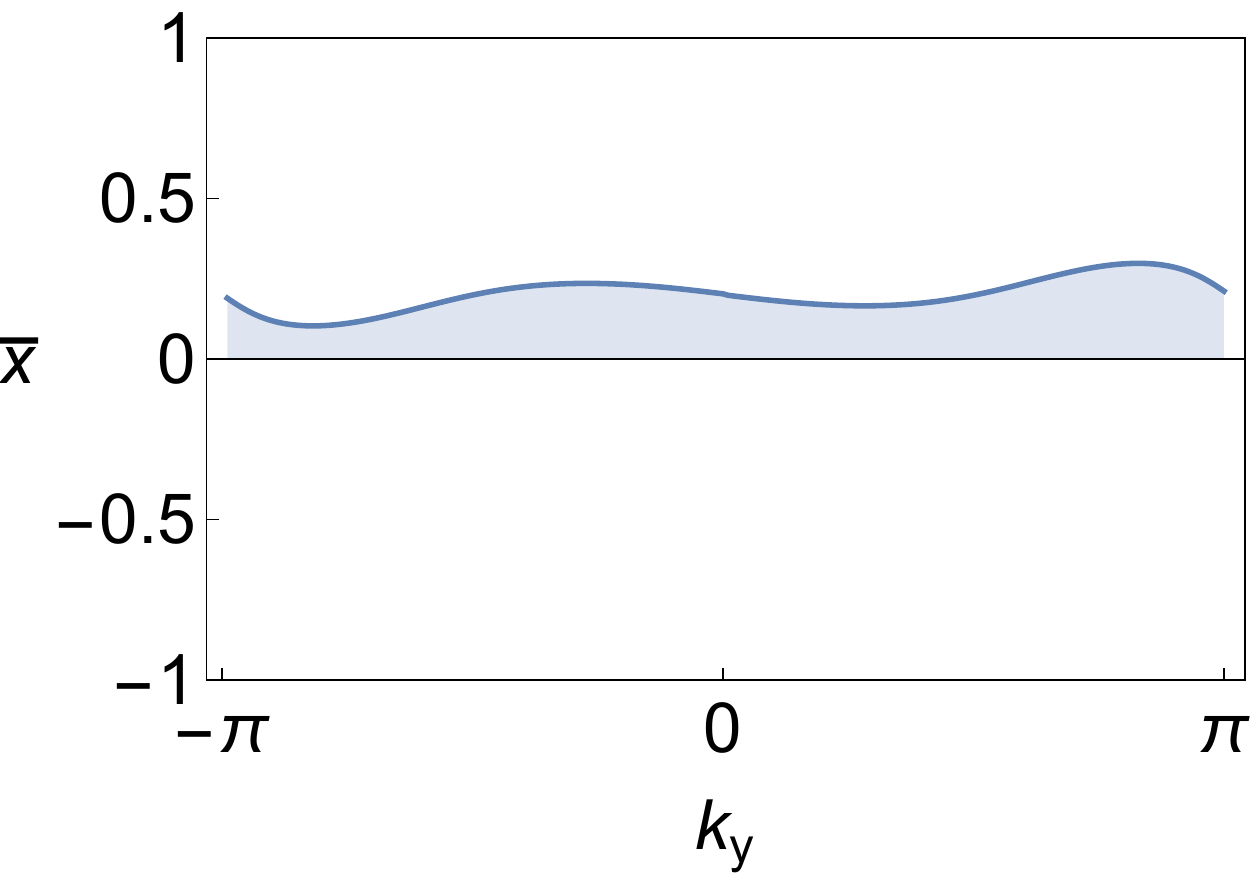}
\caption{Position of the WCC for $\mu=0, \Delta\mu=3, t=1, t'_x=1, t'_y=0.5, \alpha=2$ and $m=1,3,5,7$ The jump in the WCC positions between $-\pi$ and $\pi$ is equal to (in absolute values) $1,2,1$ and $0$ respectively.We thus get a topological phase for $m=1$ and $5$ and the trivial phase for $m=3$ and $7$.}
\label{fig.WCC}
\end{figure*}

{\it ii) Phase diagram --}
We use a numerical method inspired by the Ref.~\cite{Soluyanov_Vanderbilt_2012} 
and adapted to the AF case in Ref.~\cite{Nous} to compute the topological invariant 
of the system for several sets of parameters. This method comprises two parts. 
First we obtain a smooth definition of the eigenstates over the BZ using a parallel 
transport method, and then compute the position of the Wannier charge centers 
(WCC) as a function of $k_y$. The sum of the WCC positions over the filled band 
may not have the same value for $k_y=-\pi$ and $k_y=\pi$, but the difference of these 
two values is an integer~\cite{Vanderbilt92,Yu11,Soluyanov_Vanderbilt_2012}. 
If this integer is odd, the system is in a topological phase; if it is even, the phase 
is topologically trivial. Keeping all the other parameters fixed, we vary the strength 
$m$ of the staggered magnetization. In the Fig.(\ref{fig.EqDisp}), we see that for 
different values of $m$ the gap closes at a different TRIM -- and that each time 
the topology of the phase changes (see Fig.(\ref{fig.WCC})). One may also note 
that for $m\simeq 2.25$ the gap closes at non-TRIM points, but this does not 
change the value of the $Z_2$ invariant.

We present the obtained phase diagram in the Fig.(\ref{fig.DiagPhase}), 
and compare it to the one expected from the criterion due to Fang {\it et al.}. 
The discrepancies appear when a band inversion occurs at 
an  A-TRIM, excluded from the Fang criterion, 
such as for $m=2$ and $m=4$. We verified the above result using the 
``reconnection phase'' method, described in the Ref.~\cite{Nous}. 
These results are not presented here for brevity, but are in complete 
agreement with the WCC computation.

The Hamiltonian we discuss is block-diagonal, and can be separated 
into two blocks that are related by $\ta$-symmetry. Hence it is possible 
to work with a single block, and compute the Chern number of the 
different bands. We restrict ourselves to a stable subspace spanned by the four states 
$(\ket{\uparrow~sA},\ket{\uparrow~sB},\ket{\downarrow~pA},\ket{\downarrow~pB})$. 
We compute analytically the eigenstates as a function of $k$. Then, using the Eqs.(9) 
and (10) of the Ref~\cite{Berry}, we integrate the Berry curvature numerically to 
obtain the first Chern number. We finally sum over all the filled bands. The resulting 
phase diagram is again in perfect agreement with the WCC and the reconnection 
phase computation. 
Concerning the validity and coherence of our results, we note that the computation 
of the Berry curvature is analytical, and thus, the only error could come from 
numerical integration to obtain the Chern number. The accuracy of our 
numerical integration is well controlled and rules out an inconsistency.

Finally, we realized an explicit construction of the edge states, following the 
methods discussed in [\onlinecite{Konig08}] and [\onlinecite{Nous}]. In order 
to simplify the problem, we chose to work with a unit cell containing four sites 
(forming a square) rather than two. Despite the fact that, with this choice, we 
have to work with a 16-band model, we now have hopping only between 
nearest-neighbor unit cells, which simplified finding the edge states. 
For the same set of parameters as before, we looked for edge states 
on an antiferromagnetic edge (alternating up and down magnetization),
at the energy $E=0$. For $m=1$ and $m=5$, we found a single pair of 
edge states, while we found two pairs for $m=3$ and none at all for $m=7$. 
The parity of the number of pairs of edge states is thus once again in perfect 
agreement with the phase diagram we found (see Fig.\ref{fig.DiagPhase}).

To conclude, in this work we shed new light on topological phase transitions 
in centrosymmetric two-dimensional antiferromagnets. For such systems, 
one would like to find an easily computable form of the topological invariant such 
as in the Eq. (\ref{eq.inv_topo_xi}). Fang and co-authors proposed such a form, 
but it holds only in the presence of a symmetry that assures identical behavior 
at both of the A-TRIM, as does the $C_4$ symmetry. Without the latter, we do not 
yet have a simple expression for the topological invariant in an antiferromagnetic 
insulating phase. However, we put forward a set of numerical methods that allow 
one to capture the topological behavior of the system. We verified these results 
by direct computation of the Chern number and by the explicit construction of 
edge states. Notice that such numerical methods (WCC and ``reconnection phase'') 
are applicable even when direct computation of the Chern number is not easily 
accessible, for example when the Hamiltonian cannot be block-diagonalized by 
a fixed change of basis, as above. Finally, notice that for a three-dimensional 
antiferromagnetic insulator, the Refs. \cite{MongEssinMoore, Fang_Gilbert_Bernevig} 
proposed a topological index involving the B-TRIM only. It would be interesting 
to verify whether in three dimensions the absence of symmetry between the 
A-TRIM could affect this result as it does in two dimensions. 

It is our pleasure to acknowledge discussions with Alexey A. Soluyanov, 
whose suggestions helped to greatly improve this work. 

%
%


\end{document}